\documentclass{aa}

\usepackage{graphics}

\def\Hbeta{{\rm H}\beta}
\def\FeII{{\rm FeII}}
\def\kms{{\rm km~s^{-1}}}

\def\iras{IRAS 10026+4347}
\def\xindex{3.2}

\begin{document}
   \thesaurus{         
	1 
	(
11.02.1 
11.09.2 
11.17.2 
11.14.1 
11.19.1 
)
             }
   \title{Ultraluminous IRAS galaxy 10026+4347}

   \author{X.-Y. Xia\inst{1,4}, S. Mao\inst{2,4},
           H. Wu\inst{3,4}, Z. Zheng\inst{3,4}, Th. Boller$^{5}$,
	Z.-G. Deng\inst{6,4}, Z.-L Zou\inst{3,4}
           }
\offprints{X.-Y. Xia \\ $^\dag$ BAC is jointly sponsored by the Chinese 
Academy of Sciences and Peking University}
   \institute{
 {$^{1}$} Dept. of Physics, Tianjin Normal University, 300074
           Tianjin, P. R. China\\
 {$^{2}$} Max-Planck-Institut f\"ur Astrophysik
          Karl-Schwarzschchild-Strasse 1, 85740 Garching, Germany \\
 {$^{3}$} Beijing Astronomical Observatory, Chinese Academy of Sciences,
          100080 Beijing, P. R. China\\
 {$^{4}$} Beijing Astrophysics Center (BAC)$^\dag$, 100871
	Beijing, P. R. China\\
 {$^{5}$} Max-Planck-Institut f\"ur Extraterrestrische Physik, 
	Karl-Schwarzschild-Strasse 1, 85740 Garching bei M\"unchen, Germany\\
 {$^{6}$} Dept. of Physics, Graduate School, Chinese Academy of Sciences,
          100039 Beijing, P. R. China
 }
\authorrunning{Xia et al.} 

\date{Received ......, 1998; Accepted ......, 1998}

\maketitle

\begin{abstract}
We study optical and X-ray properties
of the ultraluminous IRAS galaxy 10026+4347. This galaxy is a
narrow-line QSO with very strong FeII emission. 
Three optical spectra were taken
over two years. The full widths at half maximum (FWHMs)
of the emission lines are constant whereas
the third spectrum seems to show a continuum change.
Intermediate-band photometry also shows a small (0.1 mag) but
significant decrease in flux. HST WFPC2 images suggest that
this object is a post-merger galaxy.
The source is X-ray luminous ($L_X \approx
10^{45} {\rm erg~s^{-1}}$) with a very soft X-ray spectrum (photon
index $\approx \xindex$). The X-ray luminosity exhibits variabilities of 
a factor of $\sim 8$ over four years and a factor of two within two days. 
During these X-ray flux changes, the X-ray
spectral shapes are consistent with no 
variation. All the optical and X-ray properties resemble
those of narrow-line Seyfert 1 galaxies (NLS1), except that
the FWHM of $\Hbeta$ is about 2500 $\kms$, larger
than that for most NLS1s. We discuss the implications of our results
on models of NLS1s.
\end{abstract}

\section{Introduction}

One of the most spectacular results of the IRAS satellite survey
was the discovery of a population of 
ultraluminous IRAS galaxies (ULIGs, see 
Sanders \& Mirabel 1996 for a review). Most ULIGs are strong interacting
and/or merging galaxies; some of them are possible post-merger galaxies
(e.g., Wu et al. 1998). 
Analysis of the spectral properties of ULIGs reveals that
about 10\% of ULIGs are Seyfert 1 galaxies (including QSOs)
(Lawrence et al. 1998; Wu et al. 1998). 
These infrared-selected Seyfert 1 galaxies have 
different properties from the optically-selected Seyfert 1s
(Lipari 1994). For example,
most of IRAS Seyfert 1s are strong or extremely strong 
optical FeII emitters and vice versa (Lawrence et al. 1997).
Most infrared-selected Seyfert 1s have relatively low soft X-ray
luminosities, very low $L_X/L_{\rm IR}$ ratios when compared with 
optically-selected ones.
The differences between the optically-selected
and infrared-selected Seyfert 1s are not yet fully understood.

There is a related class of the
so-called narrow line Seyfert 1 galaxies (hereafter NLS1, 
Osterbrock \& Pogge 1985). These galaxies are
defined by their optical emission line
properties. They have narrow hydrogen Balmer lines with typical full width
at half maximum (FWHM) $\approx$ 500-2000 $\kms$. The ratio of
[OIII]$\lambda$5007/H$\beta$ is less than 3 and most of them
have strong FeII emissions. NLS1s also show systematically 
steeper slopes in their soft X-ray continuum than normal Seyfert 1s. Some 
NLS1s exhibit rapid X-ray variabilities as well (Boller et al. 1996).

In this paper, we study the ultraluminous IRAS galaxy 10026+4347,
a narrow-line QSO selected from the QDOT redshift survey.
We show that this source is an unusual
object: it has some common characteristics with NLS1s, such as a
very soft X-ray spectrum, strong FeII emissions and rapid time
variabilities. On the other hand, it has a larger FWHM ($\sim
2500 \kms$) in the H$\beta$ line than most NLS1s. The
optical and X-ray properties of this source are presented in \S 2 and 3. 
In section 4, we discuss the nature of this object. 
Throughout this paper we use a Hubble constant of
$H_0=50~\kms~{\rm Mpc}^{-1}$ and $\Omega_0=1$.

\section{Photometric and Spectroscopic Properties}

IRAS 10026+4347 (RA=10:05:43.5, DEC=43:32:33.2 in J2000)
is a narrow-line QSO discovered in the QDOT
redshift survey with infrared luminosity $L_{IR} = 1.77\times 10^{12} L_\odot$
(Rowan-Robinson et al. 1990). This galaxy has a redshift of
$z=0.179$ (as determined from our spectra shown in Fig. 2).
At this redshift, $1^{''}$ corresponds to 4.6 kpc. 
600s exposure images were obtained with intermediate-band 
filters using the 60cm/90cm Schmidt telescope equipped with a 2048x2048 CCD
at the Beijing Astronomical Observatory (BAO). The galaxy
appears to be a point source on these images. HST snapshot images (Fig. 1)
obtained on May 1, 1997, however, clearly shows two faint close companions
within $10^{''}$ of the galaxy; these two objects are very probably
physically associated with \iras. In addition,
there are also a number of faint objects within a projected
distance of 300 kpc of \iras. It is
however unclear whether these objects are physically related to
\iras\ or just foreground or background objects.
It is also possible that
\iras\, is a post-merger galaxy at the center of a group of galaxies 
because the picture resembles the JHK images of some low redshift quasars
(Hutchings \& Neff 1997). To test this possibility,
we analyze the surface brightness profile for \iras\
using the HST snapshot image. We find that \iras\,
is well described by the
de Vaucouleurs ($r^{1/4}$) profile out to 10 kpc
(Zheng et al. 1998), similar
to other elliptical-like ULIGs such as Arp 220 and NGC 6240. The surface
profile of \iras\, thus suggests that it is a post-merger galaxy.
The apparent B and R magnitudes for this source
are about 16.2 and 16.1, respectively, from the USNO-A1.0 catalog 
(Monet 1996). With $B-R=0.1$, 
this galaxy is one of the bluest active galaxies
selected from ULIGs.
The absolute magnitudes can be calculated as
$M_B=-24.0$ and $M_R=-24.1$. The optical luminosity of this
object therefore falls into the regime of quasars.

\begin{figure}
\resizebox{\hsize}{!}{\includegraphics{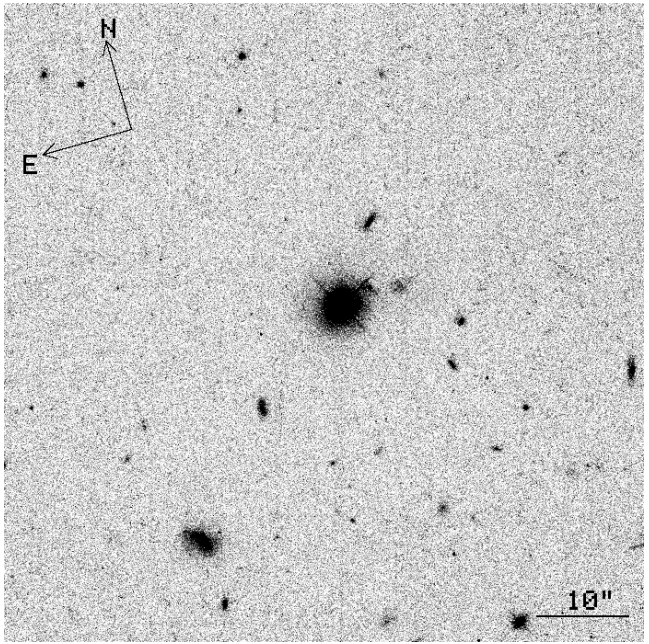}}
\caption{
HST image centered on \iras. Notice that there are two close companions
to the right of the galaxy. There are a number of faint objects
within the field. The frame is about 1.2 arcminutes on a side. The
north and east directions are indicated.
}
\end{figure}

\begin{figure}
\resizebox{\hsize}{!}{\includegraphics{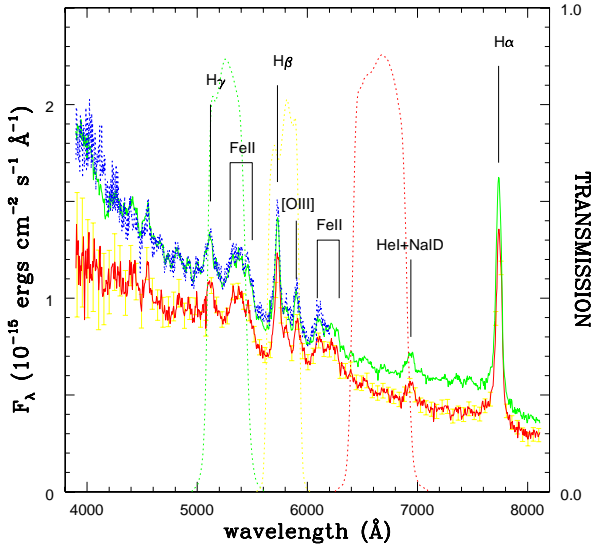}}
\caption{
Three optical spectra for \iras ~obtained in Nov. 1996 (dotted), Jan. 1997
(upper solid) and May 1998 (lowest curve). The two earlier spectra are
nearly identical and overlap with each other.
The error bars for the third spectrum are shown while those for the
two earlier curves are a factor of three smaller. The three
intermediate-band filter response curves are indicated.
Prominent emission lines are also labelled.
}
\end{figure}

The optical spectrum for this object was first obtained during
the QDOT redshift survey. It is clear from the
low dispersion spectrum that \iras~is a strong FeII emitter. We
observed this object (with higher resolution) again on Nov. 17, 1996,
Jan. 7, 1997 and May 26, 1998 using a Zeiss universal spectrograph mounted on 
the 2.16m telescope at the Xinglong Station of 
BAO. A Tektronix 1024$\times$1024 
CCD was used giving a wavelength coverage of 3500\AA\ to 6200\AA\ 
with a grating of 100\AA/mm (for Nov. 1996)
and of 3500\AA\ to 8100\AA\ with a grating 
of 200\AA/mm (for Jan. 1997 and May 1998). The spectral resolutions
are 4.7\AA\ and 9.3\AA\ (2 pixels), respectively. 
Wavelength calibration was carried out using an Fe-He-Ar lamp; 
the resulting wavelength accuracy is better than 1\AA. 
KPNO standard stars were observed to perform flux calibrations.
The three spectra are shown in Fig. 2. The absolute 
flux errors for the first two
spectra are about 5\% whereas that for the last spectrum is 
larger, around 15\%, since it was taken on a non-photometric
night. The first two spectra almost overlap with each other while 
the third spectrum seems to have a lower amplitude and a 
flatter continuum particularly at the blue end ($\lambda < 4500$\AA).
This amplitude decrease is supported by two runs of 
photometric observations in three
intermediate-bands conducted on the 60/90cm Schmidt telescope
on Feb. 21-22 and April 24-25 1998.
These three filters are 
centered on 5270\AA, 5795\AA\, and 6660\AA, with width 340\AA, 310\AA\, and
480\AA\ (cf. Fig. 2; see Fan et al. 1996 for details).
The (relative) photometry is calibrated using many stars within the 
field. and has an accuracy of about 0.03 mag.
The results indicate that there is a gradual 0.1 magnitude drop
during these runs in all three filters. Since our spectroscopic
observations indicate the emission lines are fairly stable 
(see below), the changes are therefore
from the continuum. Hence, both the photometric and
spectroscopic observations strongly suggest that
there was a small but real continuum variability.

In addition to the continuum, we have also measured the line fluxes.
The measured values
$\FeII(37,38)\lambda4750$, H$\alpha$, and $\Hbeta$ emission lines and the
FWHM for $\Hbeta$ and the flux ratio of $\Hbeta$ to FeII
are listed in table 1.
The FWHM of $\Hbeta$ is about 2500 $\kms$,
broader than the typical values ($500-2000~ \kms$) for NLS1s. 
The observed $\FeII\lambda4750/\Hbeta$ is about 2, and
since the blend of FeII $\lambda$4750 accounts for about 25\%
of the total optical FeII emission
(Collin-Souffrin et al. 1986), we infer that 
the ratio of {\it total} optical FeII to $\Hbeta$ is about 8.
Therefore, IRAS 10026+4347 qualifies as a very strong FeII emitter.
As can be seen from Table 1, 
that there is no statistically significant 
line variability during the three observations.

\begin{table}[ht]
\caption[]{
Spectroscopic Properties of IRAS 10026+4347. All fluxes are in units
of $10^{-14} {\rm erg~cm^{-2} s^{-1}}$. The FeII line refers to
the transition (37, 38) and is at $\lambda=4750$ \AA. The {\it total}
FeII to H$\beta$ ratio is roughly a factor of 4 higher. The H$\beta$
FWHMs are in units of ${\rm km ~s^{-1}}$.
}
\begin{flushleft}
\renewcommand{\arraystretch}{1}
\tabcolsep 0.11cm
\begin{tabular}{clllll}
\hline\noalign{\smallskip}
Date  & FeII & $\Hbeta$ & H$\alpha$ & FWHM & FeII/$\Hbeta$ \\
\noalign{\smallskip}
\hline
\noalign{\smallskip}
961117 &  $6.6 \pm 0.4$ & $3.3\pm 0.2$ &---& $2545\pm 60$  & 2.0 \\
970107 &  $6.6 \pm 0.4$ & $3.4\pm 0.2$ &$7.8\pm 0.4$ & $2426\pm 100$ & 1.94 \\
980526 &  $6.4 \pm 0.4$ & $3.3\pm 0.2$ &$7.3\pm 0.3$& $2638\pm 100$  & 1.94\\ 
\hline
\end{tabular}
\renewcommand{\arraystretch}{1}
\end{flushleft}
\end{table}

\vspace{-0.75cm}
\section{Soft X-ray Properties}

There are three data sets of ROSAT observation for IRAS 10026+4347 in
the ROSAT archive: the source was first
detected in the ROSAT All Sky Survey (RASS, Tr\"umper 1983; Bade et al.
1995) and later observed using ROSAT PSPC 
in May 1994 and ROSAT HRI in Nov. 1995. 
We have reduced the X-ray data using the EXSAS software at MPE
(Zimmermann et al. 1992). 
Table 2 lists the exposure time, count rate, flux and the X-ray luminosity.
Note that the flux and luminosity are model-dependent, and we have used a
power-law fit in deriving these quantities (see below).

\begin{table}[ht]
\caption[]{
Properties of IRAS 10026+4347 in the soft X-ray band (0.1-2.4 keV).
All fluxes are in units of $10^{-11} {\rm erg~cm^{-2} s^{-1}}$ 
and luminosities (last column) are in units of $10^{45} {\rm erg~
s^{-1}}$. Note that the fluxes and luminosities are derived
from a power-law fit to the spectrum.
}
\begin{flushleft}
\renewcommand{\arraystretch}{1.2}
\tabcolsep 0.11cm
\begin{tabular}{lrrcll}
\hline\noalign{\smallskip}
Instru. & date & exp. (s) & count rate & flux & $L_{\rm X}$ \\
\noalign{\smallskip}
\hline
\noalign{\smallskip}
RASS & 1990    & 508 & $0.668\pm 0.039$ &2.28    &2.4 \\
PSPC & 5/1994  & 949 & $0.085\pm 0.073$ &0.29    &0.31\\
HRI  & 11/1995 & 4711& $0.052\pm 0.019$ &0.80    &0.84\\
\hline
\end{tabular}
\renewcommand{\arraystretch}{1}
\end{flushleft}
\end{table}

  From table 2 it is obvious that the source varied during the three
observations: the count rate in RASS is about 8 times larger than
the PSPC observation conducted after 4 years.
We also performed time variability test using the RASS data.
We bin the data in an interval of 4000s. Notice that
the ROSAT wobbling is on the scale of $\sim 400$s, and so has no
effect on our variability test. We find that
the count rates exhibit about a factor 
of two variability within two days. An examination of the HRI data
indicates similar behaviors. So 
IRAS 10026+4347 is a variable object in the soft X-ray on both short and
long time scales. 

\begin{figure}
\vspace{-3.5cm}
\resizebox{\hsize}{!}{\includegraphics{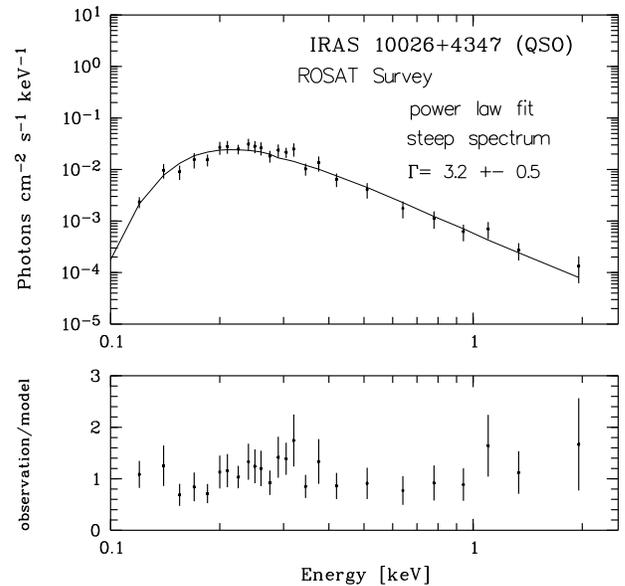}}
\caption{
Soft X-ray spectrum for \iras. The RASS data are indicated by crosses
whereas the solid line shows a power-law fit. The ratios
of the observations to model predictions are shown at
the bottom panel. Notice there are few photons above 1 keV, implying
a soft spectrum.
}
\end{figure}

We analyze the soft X-ray spectrum using the RASS data
since it has higher photon number counts than
that of the later PSPC observation.
Fig. 3 shows the spectrum together with a power-law fit.
The photon index is found to be $3.2\pm 0.5$. This value
is steeper than the average slope
of 2.5 for most AGNs, but is in the range of
NLS1s (Boller et al. 1996). 
The spectral shape from the PSPC observation is uncertain (due to limited
number of photons), but is in agreement with
that found from the RASS data. The data is therefore consistent with
no soft X-ray spectral variations while the flux has varied by a
factor of $\sim 8$.

\section{Discussion}

It is clear from optical and soft X-ray observations that \iras~ is
(like most ULIGs) a merging galaxy with
extremely strong FeII emission; in the X-ray, its luminosity is
comparable to normal Seyfert 1 galaxies or QSOs and has
steep X-ray continuum slope and
large soft X-ray variability. In addition, it exhibits 
optical continuum variability.
These optical spectral and photometric variabilities are not unusual
for normal Seyfert 1 galaxies (e.g., NGC5548, NGC4151).
IRAS 10026+4347 seems remarkably similar to another ultraluminous
IRAS galaxy RX J0947.0+4721 (Molthagen et al. 1998). Both sources are
X-ray bright with luminosities of $10^{45} {\rm erg~s^{-1}}$ and steep
soft photon index (4.2 for RX J0947.0+4721). Both objects
show large variations in the X-ray count rate 
while the soft X-ray spectrum remains stable. In the optical, 
these two sources have quasar level luminosity
($M_B=-24.7$ for RX J0947.0+4721)
and strong FeII emissions. However, these objects do have a
notable difference: 
RX J0947.0+4721 has very narrow Balmer 
lines (FWHM $\approx 1370\pm 170~ \kms$). 
The similarity between the properties of \iras\ with RX
J0947.0+4721, and more generally with the NLS1 class, is striking
despite of their differences in the $\Hbeta$ FWHM. This
strongly suggests that NLS1, narrow-line QSOs, normal Seyfert 1 galaxies
and normal QSOs form a
continuum in their properties such as $\Hbeta$ FWHM (cf. Boller et al. 1996).

Many models have been suggested to explain the puzzling characters
of NLS1s (see Boller et al. 1996 and references therein). A promising
suggestion is that NLS1s are supermassive analogues of galactic black hole
candidates in a high supersoft state; in this state, the source is
accreting close to the Eddington limit. The resulting
X-ray spectrum has a steep slope and is relatively stable (Pounds et al. 1995;
Comastri et al. 1998; Puchnarewicz et al. 1998; Wang et al. 1998).
The intense soft X-ray
flux may prevent the formation of broad-line clouds close to the central
source (see Guilbert, Fabian \& McCray 1983; White, Fabian \& Mushotzky
1984). This scenario is particularly attractive for merging galaxies
such as \iras~ (cf. Fig. 1). Numerical simulations (e.g., 
Lin et al. 1988; Mihos \& Hernquist 1996) show that large
amount of gas flows into the central regions of merging galaxies, which
can presumably fuel the black hole at the Eddington accretion
rate. The merging process may also induce intense star formations; the
resulting supernovae may enrich metals including iron
in the surrounding gas. The intense FeII emissions can therefore also be
explained. The optical and X-ray properties of \iras\ are
broadly consistent with this picture. Clearly more quantitative modelling
is needed to see whether this picture can explain the intense
variations seen in NLS1s such as 13224-3809 and PHL 1092, which
are also very luminous or ultraluminous IRAS galaxies
(Boller et al. 1996).

Galaxies like \iras\ are the best laboratory for 
studying the emission line region in AGNs. With their rapid X-ray
variability, simultaneous optical and X-ray reverberation mapping
(e.g., Krolik et al. 1991; Peterson et al. 1998)
of NLS1s will provide valuable information
on the differences between the emission mechanisms
of NLS1s and the general population of
Seyfert 1 and 2 galaxies. 

\begin{acknowledgements}

We thank Prof. J. Tr\"umper and Prof. L.Z. Fang for stimulating the project
of identification of ULIGs with RASS during which the object \iras\ was
discovered.  We thank Drs. F. Meyers and T. Wang 
for discussions and Thomas Erben for reducing the HST WFPC2
images. We are also indebted to the BATC group members of BAO, especially to 
Dr. Jin Zhu, for performing the photometric observations of \iras.
This project was partially supported by the NSF of China 
and NSFC-DSF exchange program.
\end{acknowledgements}

\end{document}